# Reversibility Window, Aging, and Nanoscale Phase Separation in $Ge_xAs_xS_{1-2x}$ Bulk Alloy Glasses

*Tao Qu and P. Boolchand*
Department of Electrical Computer Engineering and Computer Science
University of Cincinnati, Cincinnati, Ohio 45221-0030

The non-reversing enthalpy ($\Delta H_{nr}$) near $T_g$, in bulk $Ge_xAs_xS_{1-2x}$ glasses is found to display a global minimum (~0) in the 0.11 < x < 0.15 range, the *reversibility window*. Furthermore, the $\Delta H_{nr}$ term is found to age for glass compositions below ( x < 0.11) and above ( x > 0.15) the window but *not* in the window. In analogy to corresponding selenides, glass compositions in the *window* represent the *intermediate phase*, those at x < 0.11 are *floppy*, and those at x > 0.15 *stressed-rigid*. Raman scattering shows *floppy* and *stressed rigid* networks to be partially nanoscale phase separated, an aspect of structure that contributes to a narrowing of the *reversibility window width* and to suppression of the $\Delta H_{nr}$ term in S-rich glasses qualitatively in relation to corresponding Se-rich glasses.

Ideas based on Lagrangian bonding constraints have proved to be central to understand the *elastic behavior* of network glasses[1,2]. In the 1980s, these considerations led to the recognition[3] that a *random* network of *floppy* chains would spontaneously become *rigid* when the cross-link density or mean coordination number, *r*, exceeds a threshold value of 2.40. Stimulated by Raman scattering results[4-8] on chalcogenide glasses, recent numerical simulations[9] and combinatorial calculations[10] have shown that the rigidity transition ($r_c(1)$) will, in general, precede the stress transition ($r_c(2)$) in *self-organized* networks. In the intervening region, $r_c(1) < r < r_c(2)$, the backbone is viewed to comprise of *select local structures* that are rigid but stress-free. A count of Lagrangian constraints (due to bond-stretching and bond-bending constraints) shows that these local structures possess 3 constraints per atom ($n_c$), equal to the 3 degrees of freedom an atom has in a 3d structure. The match insures the stress-free character (isostatically rigid) of the backbone. Networks possessing[9-12] $r < r_c(1)$ are viewed to be *floppy* (underconstrained, $n_c < 3$), those in the intervening region ($r_c(1) < r < r_c(2)$) to be *intermediate* (optimally constrained, $n_c = 3$), and those at $r > r_c(2)$ to be *stressed-rigid* (overconstrained, $n_c > 3$).

Calorimetric results on chalcogenide glasses have independently provided thermal fingerprints[5-8,11-16] of the three elastic phases, viz., *floppy, intermediate and stressed-rigid*. Specifically, the *non-reversing enthalpy* ($\Delta H_{nr}$) near $T_g$ deduced from T-modulated DSC measurements (MDSC), is found to nearly vanish (*thermally reversing windows*) in *intermediate phases*, and to be usually an order of magnitude larger in both the *floppy* and *stressed-rigid* phases. Glass transitions in *intermediate phases* are thus found to be almost completely *thermally reversing* in character, and furthermore are found not to age. An illustrative example is the case of the $Ge_xAs_xSe_{1-2x}$ ternary[6,14] on which comprehensive Raman and MDSC results have independently tracked the *rigidity* ($r_c(1) =$ 2.27) and *stress* ($r_c(2) = 2.55$) transitions. Both probes independently show[6] the *intermediate phase* to extend over the $0.09 < x < 0.155$, or $2.27 < r < 2.46$ composition range, since $r = 2 + 3x$ (ref.14).

The $Ge_xAs_xSe_{1-2x}$ ternary is an ideal test system to examine these connectivity related phase transitions because the backbones are nearly fully polymerized[17]. This is, however, not the case in the present $Ge_xAs_xS_{1-2x}$ ternary. Both S-rich ($x < 0.10$) and S-deficient ($x > 2/11$) glasses, apparently *nanoscale phase separate* (NSPS). Given the similarity of local structures between these two ternaries, and the availability of rather comprehensive results on the elastic phases of the selenides, examination of the sulfide glasses provides a means to elucidate the role of NSPS effects on the underlying elastic phases. In this work we show that in the sulfide glasses, the *reversing window* is narrower ($2.33 < r < 2.45$), and backbone related *aging effects* outside this *window* are less pronounced in relation to those seen in the fully polymerized selenide counterparts.

Elemental S, $As_2S_3$, Ge and As of 99.999% purity from Cerac Inc., were used as starting materials. They were reacted in evacuated quartz ampoules of 5mm id, homogenized at 900ºC for 48 hours and equilibrated at 50ºC above the liquidus before a water quench[6,14]. Samples were allowed to relax at room temperature for 4 weeks or longer before initiating Raman and MDSC measurements. Raman scattering was excited using 647 nm radiation and the backscattered light collected at 1 cm$^{-1}$ resolution in a triple monochrometer system (Jobin Yvon T 64000) with a CCD detector. The MDSC measurements used a TA instruments model 2920 unit operated at 3ºC/min scan rate and 1ºC /100s modulation rate.

In Fig.1 we compare MDSC scan of a S-rich glass with that of pure S. Common to both scans is the S polymerization transition ( endotherm- $\Delta H_{nr}(T_\lambda)$) near $T_\lambda$ = 154º C which constitutes signature for opening of $S_8$ crowns into $S_n$-chains[18]. The shape of the non-reversing heat flow associated with the $T_\lambda$ transition is rather curious; it has a *precursive exothermic* event in the glass but not in pure S. This suggests that $S_8$ crowns relax towards the *network backbone* prior to opening into $S_n$- chains and being incorporated into the backbone at T > $T_\lambda$. The endotherm near T =72º C represents the glass transition, and $T_g$ was deduced from the inflexion point of the *reversing heat flow* signal[13-17]. The *non-reversing* heat flow associated with $T_g$ is the hash-marked peak labeled as $\Delta H_{nr}$ ($T_g$). Fig.2 provides a summary of the MDSC results; panel (a) shows $T_g(x)$ to increase with x except for a glitch near x = 0.17, panel (b) shows variations in $\Delta H_{nr}(x)$ displaying a global minimum in the 0.11 < x < 0.15 range, and panel (c) shows variations in the specific heat jump $\Delta C_p(x)$ displaying profound aging effects. Note that within the *thermally reversing window*, $\Delta H_{nr}$ term does not age, although rather striking aging effects occur for glass compositions outside this *window*.

Fig. 3a shows Raman lineshapes observed in the glasses as a function of composition x. Vibrational bands in the 300 - 400 cm$^{-1}$ bond-stretching range result from the Ge- and As-centered local structures that contribute to the network backbone as discussed earlier[6,19]. Fig. 3b shows an enlarged view of the sharp modes (boxed area) observed in the spectrum of a glass at x = 0.20. This glass composition corresponds to y = ½ in the $(As_2S_3)_y(Ge_2S_3)_{1-y}$ ternary that was recently investigated in Raman scattering and MDSC measurements by Mamedov et al[19]. The octet of modes labeled as $\nu_n$ with n going from 1 to 8 reported here are in harmony with results of ref. 19. Molar volumes ($V_m$) of the glasses were recently reported by B.Aitken et al.[20] and their results are reproduced for the reader's convenience in fig. 4a. In fig.4a, we also compare variations in $V_m$ to those in Raman mode ($\nu_6$ and $\nu_8$) scattering strengths as a function of x, and note that all these observables show a global maximum near the glass composition, x = 0.20.

In fig. 4b, we compare the *thermally reversing window* in ternary $Ge_xAs_xSe_{1-2x}$ glasses[6,14] (2.27 < $r$ < 2.46), with that observed in the present glasses (0.11 < x < 0.15, or 2.33 < $r$ < 2.45 range). By analogy, we identify the latter window as the *intermediate phase* of the present glasses. The nature of the isostatic building blocks in the two ternaries share commonality, and consist of As-centered quasi-tetrahedral(QT), $AsX(X_{1/2})_3$ and pyramidal (PYR), $As(X_{1/2})_3$ units, and Ge centered corner-sharing (CS) $GeX_4$, and edge sharing (ES) $GeX_2$ units with X = S or Se. For each of these units, $n_c$ = 3 as shown earlier[14]. The mean coordination number, $r$, of QT, PYR, CS and ES units respectively are 2.27, 2.40, 2.40 and 2.67 and the intrinsic spread of $r$ of these units serves to determine the width of the *intermediate* phase. The lower-edge of the intermediate phase is at $r_c(1)$ = 2.33 in the present sulfides (Fig.4b), somewhat upshifted in relation to the one in selenides ($r_c(1)$ = 2.27), we suppose because significant amount of S segregates into crowns in S-rich glasses and thereby inhibits growth of the backbone. The upper-edge of the intermediate phase is at $r_c(2)$ = 2.45, nearly the same as in the case of the selenides, $r_c(2) = 2.46$. Not surprisingly the $S_8$ concentration extrapolates to zero near r~ 2.40. Interestingly, a small but decreasing concentration of $S_8$ crowns persists in the *intermediate phase* of the present glasses, and apparently that does not influence the *self-organized* nature of the backbone probably because $\Delta H_{nr}$ is intrinsically a backbone property.

Some of these sharp modes labeled $S_n$ in Fig 3a and the $T_\lambda$ transition in Fig. 1a constitute signatures of $S_8$ crowns in the glasses. A detailed analysis[21] of the Raman scattering has permitted to separate contributions of $S_8$ crown modes from $S_n$-chain modes. Thus, for example, we find that the mode at 153 cm$^{-1}$ is a $S_8$ crown mode, and variations in scattering strength of this mode as a function of x correlate well with the $\Delta H_{nr}(T_\lambda)$ enthalpy, providing internal consistency to the assignments.

In the present ternary a chemical threshold ($x = x_t$) is predicted[6] to occur when all sulfur becomes bonded in As-centered and Ge centered local structures,

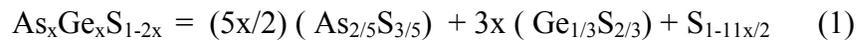

$$As_xGe_xS_{1-2x} = (5x/2)\,(As_{2/5}S_{3/5}) + 3x\,(Ge_{1/3}S_{2/3}) + S_{1-11x/2} \quad (1)$$

leaving the coefficient of the free S term in equation (1) to vanish at the threshold, $x = x_t$, i.e., $1-11x_t/2 = 0$, or $x_t = 2/11$ or 0.182. At $x > x_t$, one expects homopolar (As-As, Ge-Ge) bonds to manifest in the network. In the present ternary, As-As bonds apparently first manifest in $As_4S_4$ and $As_4S_3$ *monomers* in the $0.16 < x < 0.23$ range as revealed in Raman scattering. Of the octet of modes observed at $x = 0.20$ (Fig.3b), $\nu_6$ and $\nu_8$ come exclusively from $As_4S_4$ and $As_4S_3$ monomers[19] respectively. And the correlation in scattering strengths of these modes with molar volumes (Fig. 4a) showing a global maximum near x = 0.20, serves to confirm[19] that glasses in the 0.16 < x < 0.23 range are partially NSPS. The non-bonding van der Waals forces segregate these monomers from each other and the backbone. Here we must remember that the sulfur van der Waals radius[22] ( 180 pm) is nearly twice as large as the sulfur covalent radius ( 102 pm). NSPS leads to a loss in network packing that is reflected in the molar volumes acquiring a maximum near x = 0.20. The result is in harmony with the Raman scattering strength of the $v_6$ and $v_8$ modes, that peak near x = 0.20. These modes were identified[19] with the two monomers in question, $As_4S_4$ and $As_4S_3$ ( Fig. 3c).

Finally, the results of Fig.4b permit a quantitative comparison of aging of the $\Delta H_{nr}$ term in the sulfide glasses with that in corresponding selenides, particularly in the floppy regime. Both sets of MDSC measurements were performed 6 months after melt-quenching, with glass samples aged at room temperature below $T_g$. We note that $\Delta H_{nr}$ ~ 0.40 kJ/mole in a selenide glass at x = 0.06, but only 0.10 kJ/mole for the corresponding sulfide glass. The 4-fold reduction of $\Delta H_{nr}$ reflects the reduced polymer/monomer fraction of the sulfide glass in relation to the selenide glass .

In summary, MDSC measurements on $As_xGe_xS_{1-2x}$ glasses reveal that the *reversibility window* in the 0.11 < x < 0.15 range. Aging of the non-reversing enthalpy is observed for glass compositions outside the window but *not* in the *window*. A distinguishing feature of the present glasses is that $S_8$ crowns segregate from the backbone qualitatively in the *floppy* regime as observed both in Raman scattering and MDSC measurements. The qualitative loss of backbone suppresses aging of the non-reversing enthalpy in S-rich glasses in relation to corresponding Se-rich glasses. We have benefited from discussions with D. McDaniel, B.Goodman and S. Mamedov during the course of this work. This work is supported by NSF grant DMR-01-01808.

Figure Captions

**Fig.1**. MDSC scan of (a) x = 0.08 sulfide glass and (b) pure S taken at 3ºC/min and 1ºC/100s modulation rate. Common to both scans is the sulfur polymerization transition $T_\lambda$ near 159º C. Note that the non-reversing heat associated with $T_\lambda$ transition has a precursive exotherm in the glass but not in pure S (inset of Fig.1b). In pure S, we observe the α→β transition near 102ºC and the melting transition near 118ºC.

**Fig.2**. Summary of MDSC results on sulfide glasses showing trends in (a) glass transition temperature, $T_g(x)$, (b) non-reversing heat, $\Delta H_{nr}(x)$ and (c) specific heat jump near $T_g$, $\Delta C_p(x)$ as a function of glass composition. Note the reversibility window in the 0.11 < x < 0.15 range in panel (b).

**Fig.3**.(a) Raman scattering in sulfide glasses as a function of increasing x, (b) enlarged view of sharp modes observed in the 150 - 300 cm$^{-1}$ range, boxed area, for a sample at x = 0.20, (c) Molar volumes of the glasses taken from ref. 20 and scattering strength of Raman modes $v_6$ and $v_8$ from present work show a maximum near x = 0.20, ascribed to partial NSPS.

**Fig. 4** (a) Molar non-reversing enthalpy, $\Delta H_{nr}(x)$, in $Ge_xAs_xS_{1-2x}$ glasses (●) and $Ge_xAs_xSe_{1-2x}$ glasses (○) taken after a 6month waiting time ($t_w$). Note the smaller width of the reversibility window in the sulfides than in the selenides. Also note the much smaller enthalpy in S-rich glasses than in the Se-rich glasses because of a qualitative loss of the backbone due to NSPS in the former as x → 0.

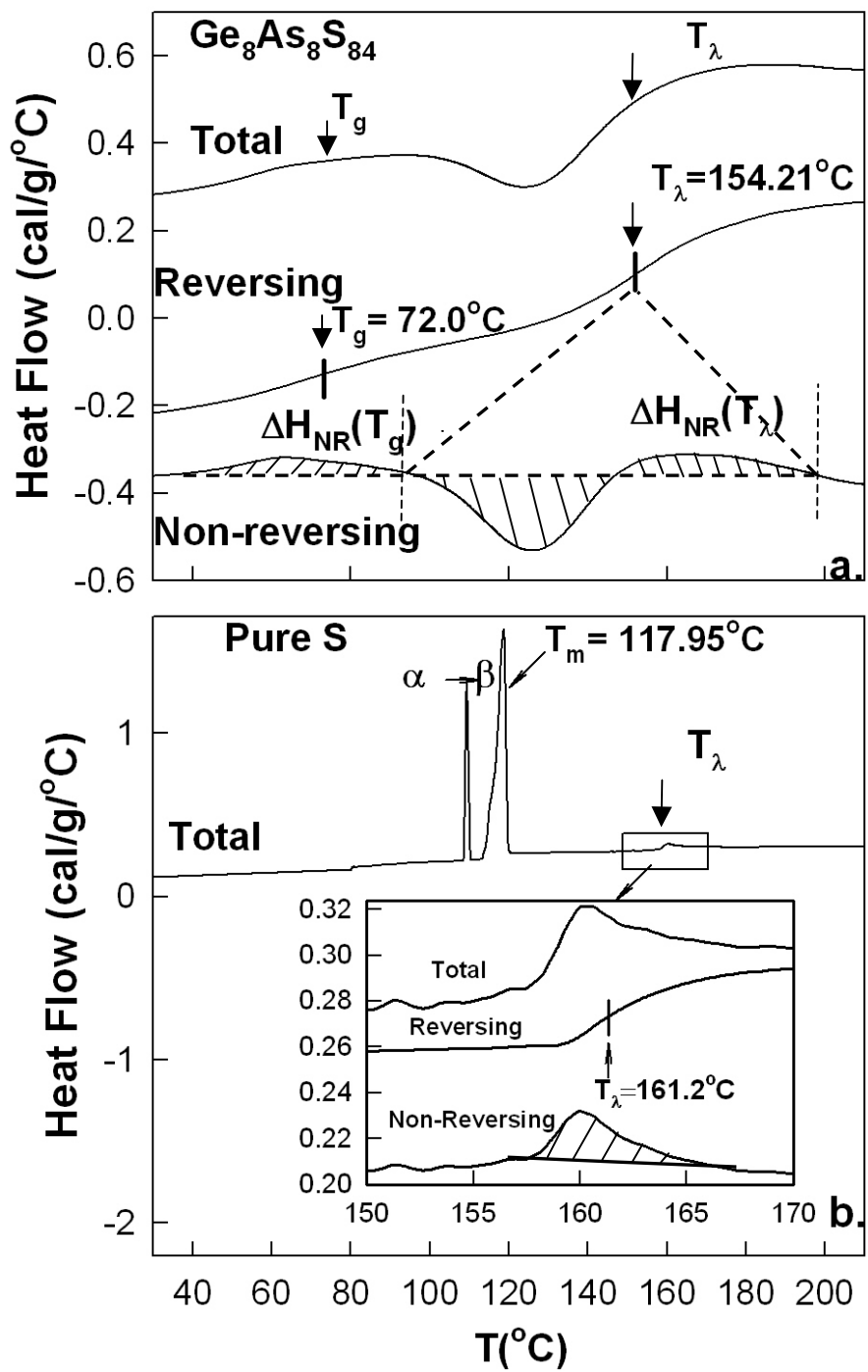

FIG 1

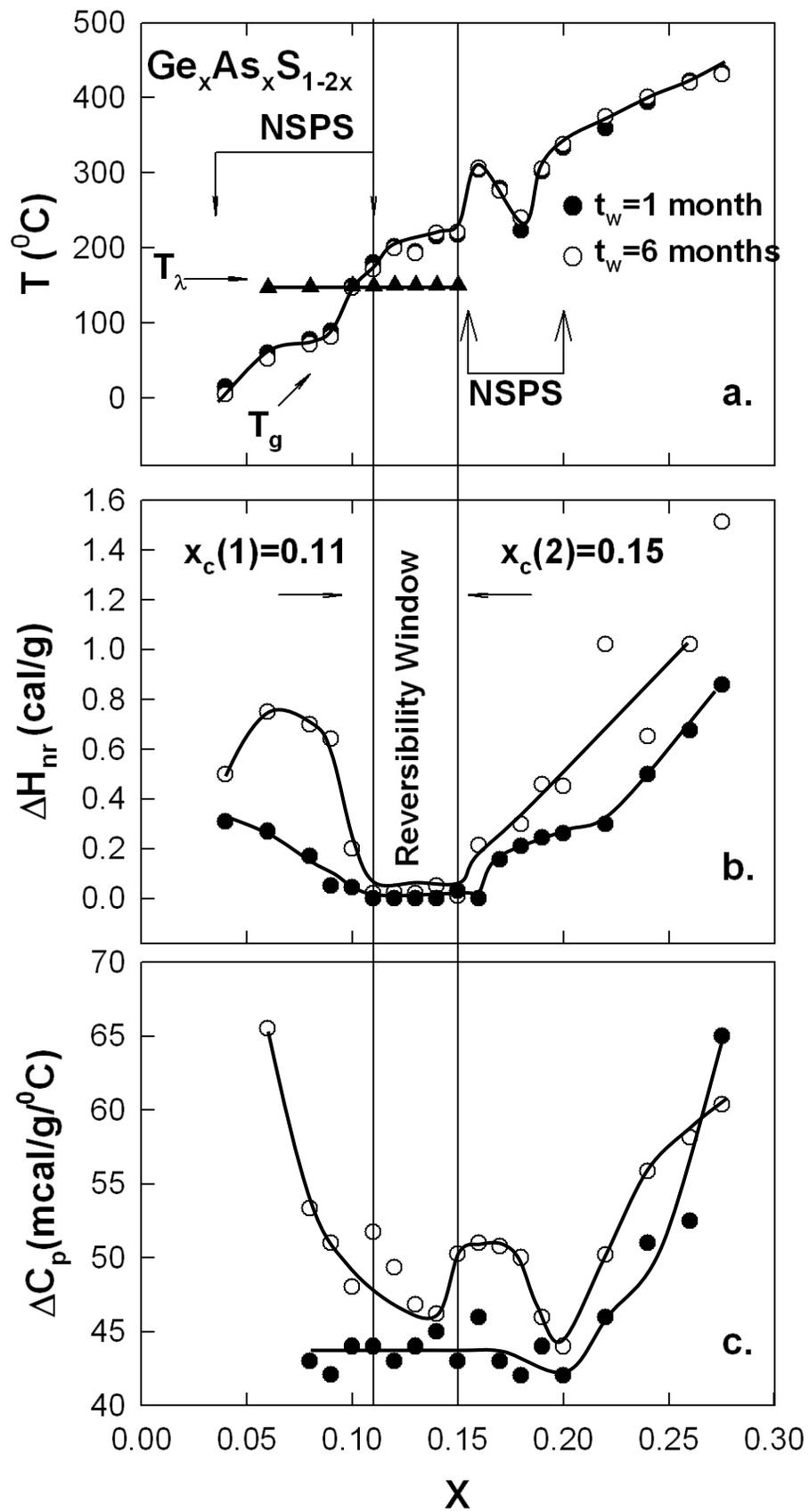

FIG 2

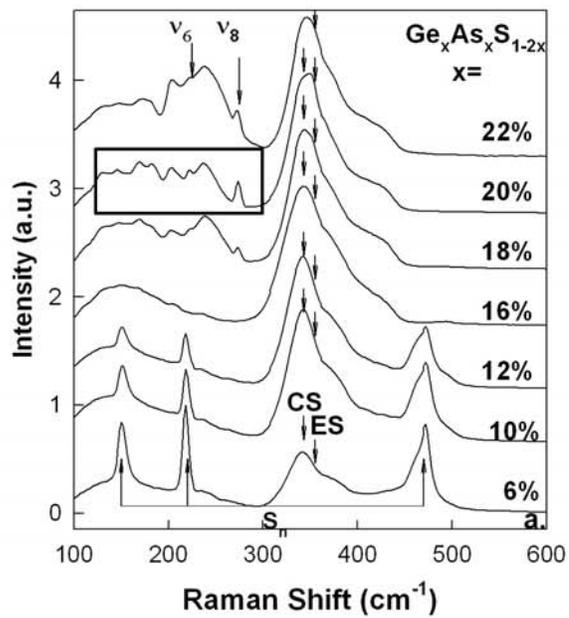
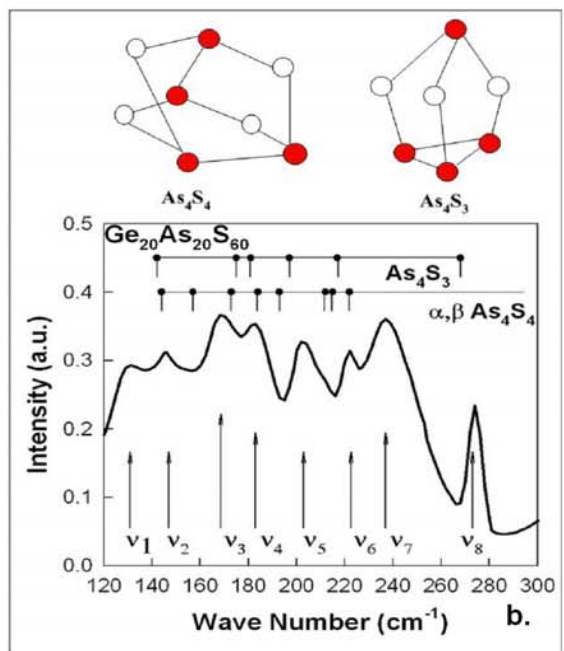
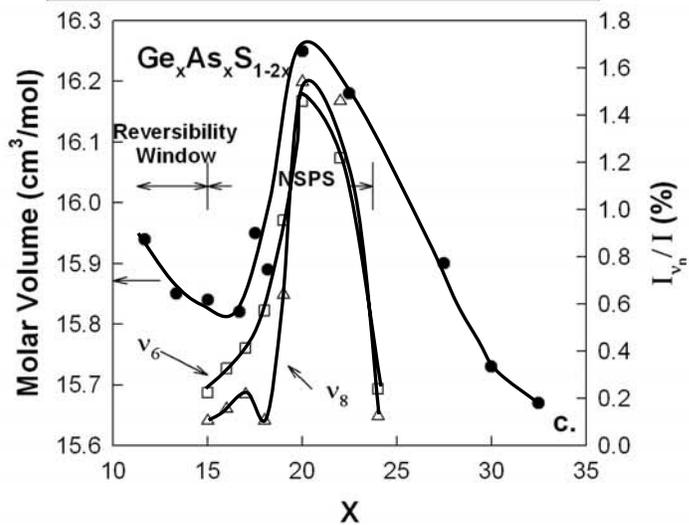

Fig 3

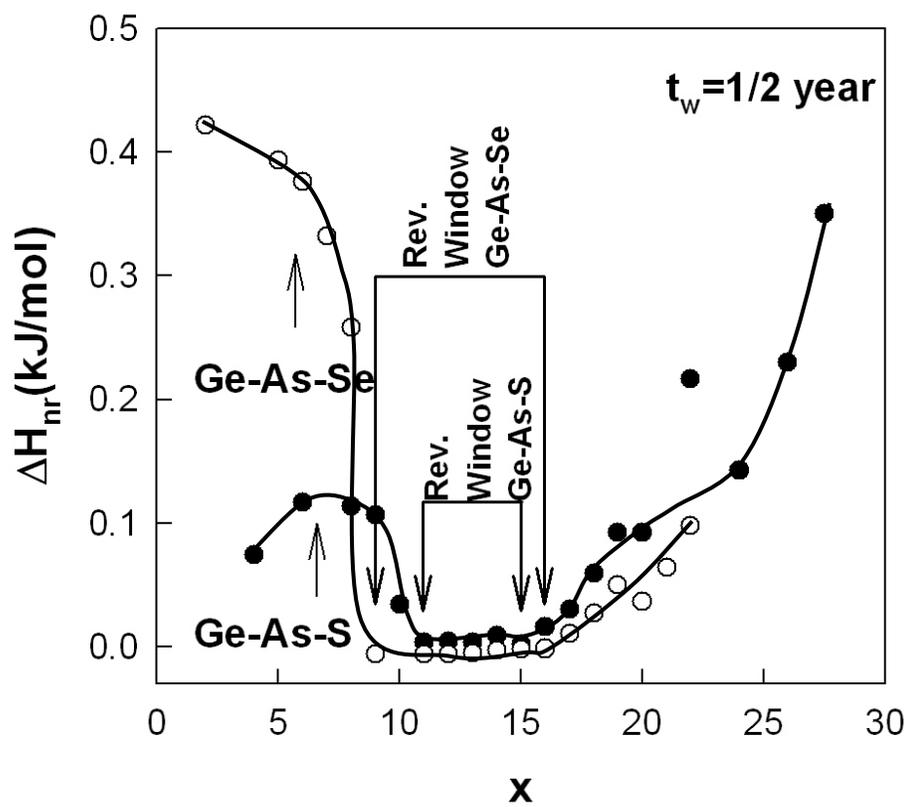

Fig 4